\documentclass[preprint,showpacs,preprintnumbers,amsmath,amssymb,nofootinbib]{revtex4}

\usepackage{etex}

\usepackage{amsthm,amscd,amsbsy,array}
\usepackage{bm}% bold math
\usepackage{soul} % underline, strikethrough, etc.

\usepackage{graphics,graphicx,xcolor}

\usepackage{soul} % underline, strikethrough, 

\usepackage[colorlinks=true, pdfstartview=FitV, linkcolor=blue, citecolor=blue, urlcolor=blue]{hyperref} % hyperref

\newcommand{\cyan}{\textcolor{cyan}}

\newcommand{\red}{\textcolor{red}}  
\newcommand{\blue}{\textcolor{blue}}

\newcommand{\gb}{\colorbox{green}}

\newcommand{\dgreen}{\textcolor[rgb]{0,0.35,0}}

\newenvironment{redtext}{\color{red}}{\ignorespacesafterend} 
\newenvironment{bluetext}{\color{blue}}{\ignorespacesafterend}

\newcommand{\bblue}{\begin{bluetext}} 
\newcommand{\eblue}{\end{bluetext}} 
\newcommand{\bred}{\begin{redtext}}
\newcommand{\ered}{\end{redtext}}
\numberwithin{equation}{section}

\let\ssection=\section
\renewcommand{\section}{\setcounter{equation}{0}\ssection}

\newcommand{\cA}{{\mathcal{A}_{+}}}

\newcommand{\bb}{{\bf b}}

\newcommand{\bc}{{\mathbf{c}}}

\newcommand{\cB}{{\mathcal{A}_{\times}}}

\newcommand{\Id}{\mathrm{Id}}

\newcommand{\bk}{\mathbf{k}}

\newcommand{\bp}{{\bf p}}

\newcommand{\bx}{{\bm{x}}}

\newcommand{\bbR}{\mathbb{R}}

\newcommand{\bX}{{\bf X}}

\def\smallover#1/#2{\hbox{$\textstyle\frac{#1}{#2}$}} %

\def\bp{{\bm{p}}}

\def\benu{\begin{enumerate}}
\def\eenu{\end{enumerate}}
\def\beq{\begin{equation}}
\def\eeq{\end{equation}}
\def\beqa{\begin{eqnarray}}
\def\eeqa{\end{eqnarray}}

\def\barray{\left(\begin{array}}
\def\earray{\end{array}\right)}
\def\barraynb{\begin{array}}
\def\earraynb{\end{array}}

\def\?{\quad{\gb{\fbox{\texttt{?}}\;}}\quad}
\def\p{{\partial}}

\def\v0{\mathbf{0}}

\def\beq{\begin{equation}}
\def\eeq{\end{equation}}
\def\bea{\begin{eqnarray}}
\def\eea{\end{eqnarray}}

\def\p{\partial}

\def \p{{\partial}}

\def\6{\partial}
\def\7{\tilde}
\def\8{\widehat}
 \def\bx{{\bf x}}

\def\G11{\Gamma_{11} }

\newcommand{\const}{\mathop{\rm const.}\nolimits}
\newcommand{\half }{\frac{1}{2}}

\def\smallover#1/#2{\hbox{$\textstyle\frac{#1}{#2}$}} %
\def\smallcirc{{\raise 0.5pt \hbox{$\scriptstyle\circ$}}}
\def\2{{\smallover1/2}}

\let\ssection=\section
\renewcommand{\section}{\setcounter{equation}{0}\ssection}

\def\besub{\begin{subequations}}
\def\esub{\end{subequations}}
\begin{document} 

\preprint{\texttt{arXiv:1803.09640v3  [gr-qc]}
}

\title{Sturm-Liouville and Carroll~: at the heart of the Memory Effect
\\[6pt]
}

\author{
P.-M. Zhang${}^{1}$\footnote{e-mail:zhpm@impcas.ac.cn},
M. Elbistan${}^{1}$\footnote{mailto:elbistan@impcas.ac.cn.},
G. W. Gibbons${}^{2}$\footnote{mailto:G.W.Gibbons@damtp.cam.ac.uk},
P. A. Horvathy${}^{1,3}$\footnote{mailto:horvathy@lmpt.univ-tours.fr}
}

\affiliation{
${}^1$Institute of Modern Physics, Chinese Academy of Sciences
\\ Lanzhou, China
\\
${}^2$D.A.M.T.P., Cambridge University, U.K.
%\\ Wilberforce Road,
%\\ Cambridge CB3 0WA, U.K.},
\\
${}^3$ Institut Denis-Poisson CNRS/UMR 7013 - Universit\'e de Tours - Universit\'e d'Orl\'eans Parc de Grammont, 37200, Tours, (France)
France
}

\date{\today}

\pacs{
04.20.-q  Classical general relativity;\\ 
02.20.Sv  Lie algebras of Lie groups;\\
04.30.-w Gravitational waves 
\\
\vskip 1mm
General Relativity and Gravitation (2018) 50:107 https://doi.org/10.1007/s10714-018-2430-0\\
\texttt{arXiv:1803.09640  [gr-qc]}
}

\begin{abstract} 
 For a plane gravitational wave whose profile is given, in Brinkmann coordinates, by
a $2\times2$ symmetric traceless matrix $K(U)$, the matrix Sturm-Liouville equation $\ddot{P}=KP$ plays a multiple and central r\^ole: (i) it determines the  isometries; (ii) it  appears as the key tool for switching from Brinkmann to BJR coordinates and vice versa; (iii) it determines the trajectories of particles initially at rest. All trajectories can  be obtained from trivial ``Carrollian'' ones by a suitable action of the (broken) Carrollian isometry group.
\end{abstract}

\maketitle

\tableofcontents

%%%%%%%%%%%%%%%%%%%%%%%%%%%%%%%%%%%%%%%%%%%%%%%%%%%%%%%%%%%%%%%%%%%%%%%%%%%%%%
%%%%%%%%%%%%%%%%%%%%%%%%%%%%%%%%%%%%%%%%%%%%%%%%%%%%%%%%%%%%%%%%%%%%%%%%%%%%%%
\section{Introduction}\label{Intro}
%%%%%%%%%%%%%%%%%%%%%%%%%%%%%%%%%%%%%%%%%%%%%%%%%%%%%%%%%%%%%%%%%%%%%%%%%%%%%%
%%%%%%%%%%%%%%%%%%%%%%%%%%%%%%%%%%%%%%%%%%%%%%%%%%%%%%%%%%%%%%%%%%%%%%%%%%%%%%

The motion of test particles under the influence of a gravitational wave (GW), called the \emph{Memory Effect} \cite{ZelPol,BraGri}, has attracted considerable attention as a potential tool to detect gravitational waves. Approximating a gravitational wave by an exact plane wave reveals, in particular, that particles initially at rest will move, after the wave has passed, with constant but non-vanishing relative velocity~: this is the \emph{Velocity Effect} \cite{EhlersKundt, Sou73,BraTho, BoPi89, GriPol,ShortLong,Lasenby,PolPer}. 

In this Letter we point out the central role played by (i) a\emph{ matrix Sturm-Liouville equation} \cite{Gibb75}  
\beq
\ddot{P}=KP\,,
\label{SLPeq}
\eeq 
where $K(U)$ is the profile of the wave, and (ii) by Carroll(type) symmetry \cite{Leblond}.
Eqn. (\ref{SLPeq}) (i) determines the isometries 
 ; (ii) appears as the key tool for switching from Brinkmann (B) to Baldwin-Jeffery-Rosen (BJR) coordinates and vice versa;  (iii) determines the trajectories of particles initially at rest.  

In BJR coordinates the symmetries and the trajectories are both conveniently determined in terms of another matrix, $H(u)$ in (\ref{Hmatrix}) below. In terms of B coordinates, this role is overtaken by the matrix $Q=PH$ in (\ref{BinB}), which satisfies again the
Sturm-Liouville equation above.
\goodbreak

Generic gravitational waves have long been known to have a 5-parameter isometry group \cite{BoPiRo,Sou73,exactsol}, recently identified as \emph{the subgroup of the Carroll group with rotations omitted}  \cite{Leblond,Carroll4GW} \footnote{The Carroll group \cite{Leblond} is the subgroup of the Bargmann group with no time translations; the latter is itself the subgroup of the Poincar\'e group which leaves $\p_V$ invariant. The Bargmann group is a $1$-parameter central extension of the Galilei group upon which it projects when translations along $V$ are factored out.  
 For circularly polarised periodic waves the symmetry can be extend to a $6$-parameter group \cite{exactsol,PolPer}.}.

Carroll symmetry  has long been considered as a mathematical curiosity irrelevant for physics, for the good reason that a \emph{particle with Carroll symmetry can not move} \cite{Leblond, Ancille, Carrollvs, Bergshoeff}. In this Letter we point out that (broken) Carroll symmetry does play a fundamental r\^ole, namely in describing  \emph{particle motion} in a gravitational wave background.

%\goodbreak
 
%%%%%%%%%%%%%%%%%%%%%%%%%%%%%%%%%%%%%%%%%%%%%%%%%%%%%%
\section{Killing vectors and isometries}
\label{PGW}
%%%%%%%%%%%%%%%%%%%%%%%%%%%%%%%%%%%%%%%%%%%%%%%%%%%%%%

%\subsection{In Brinkmann coordinates}

In Brinkmann (B) coordinates $(\bX,U,V)$
the profile of a plane gravitational wave is given by
 the symmetric and traceless $2\times2$ matrix
$K(U)=K_{ij}(U)$ \cite{Brink,BoPiRo,exactsol},
\begin{subequations}
\begin{align}
ds^2=\delta_{ij} dX^i dX^j + 2dUdV + K_{ij}(U) X^i X^j dU^2 \red{\,},
\label{Bmetric}
\\[6pt]
K_{ij}(U){X^i}{X^j}=
\half{\cA}(U)\big((X^1)^2-(X^2)^2\big)+\cB(U)\,X^1X^2\,,
\label{Bprofile}
\end{align}
\label{genBrink}
\end{subequations}
where $\cA$ and $\cB$ are the $+$ and $\times$ polarization-state amplitudes. 
Previously  we studied~: 
(i) linearly polarized waves: ${\cB}=0$ and  ${\cA}$ is typically a [derivative of a] Gaussian \cite{ShortLong,ImpMemory} or a Dirac delta \cite{PodSB,ImpMemory};
(ii) 
circularly polarized sandwich waves \footnote{A sandwich wave is one which vanishes outside some finite interval $[U_i,U_f]$ \cite{BoPiRo,BoPi89}.}   \cite{PolPer}~:
\beq
{\cA} (U) = C^2\,\frac{\lambda}{\sqrt{\pi}}\,e^{-\lambda^2U^2} \cos(\omega U)\red{\,},
\qquad 
{\cB} (U) = C^2\,\frac{\lambda}{\sqrt{\pi}}\,e^{-\lambda^2U^2}
\sin(\omega U)\red{\,};
\label{polgaussprof}
\eeq
(iii) circularly polarized waves with periodic  profile \cite{PolPer}
\beq
{\cA} (U) = C^2 \cos(\omega U)\red{\,},
\qquad
{\cB} (U) = C^2\,
\sin(\omega U)\red{\,}.
\label{PerProf}
\eeq

Throughout this Letter, we limit ourselves to sandwich waves with $C^1$ profile. Impulsive waves with Dirac delta-function profiles have been studied elsewhere \cite{ 
PodSB,ImpMemory}.

%%%%%%%%%%%%%%%%%%%%%%%%%%%%%%%%%%%%%%%%%%%%%%%%%%%%%%%%%%%%%%%%%%%%%%%%%%%%%%
%\section{Killing vectors and isometries}\label{Carrolliso}
%%%%%%%%%%%%%%%%%%%%%%%%%%%%%%%%%%%%%%%%%%%%%%%%%%%%%%%%%%%%%%%%%%%%%%%%%%%%%%

%%%%%%%%%%%%%%%%%%%%%%%%%%%%%%%%%%%%%%%%
%\subsection{In Brinkmann coordinates}\label{BCarrollSec}
%%%%%%%%%%%%%%%%%%%%%%%%%%%%%%%%%%%%%%%%

A particularly clear approach to isometries is that of Torre \cite{Torre} who pointed out that,  in Brinkmann coordinates $(\bX, U, V)$, the Killing vectors are
\beq
S_i(U)\p_i + \dot{S}_i(U) X^i\, \p_V\,,
\qquad
\p_V,
\label{TorreSymm}
\eeq
where the dot means $d/dU$, and
where $S_i,\, i=1,2 $ is a solution of the vector equation
\beq
\ddot{S}_i(U) = K_{ij}(U) S_j(U)\,.
\label{SLeqn}
\eeq

The simplest example is that of Minkowski space, $K_{ij}\equiv0$,  when (\ref{SLeqn}) is solved by $S_i=\gamma_i+\beta_iU $
 and (\ref{TorreSymm}) is a combination of translations in the transverse plane and two infinitesimal
 Galilei boost lifted to (\ref{genBrink}) viewed as a Bargmann space \cite{Bargmann,DGH91},
\beq
Y=(\gamma_i+U\beta_i)\p_i + \big(\delta + X^i\beta_i\big)\, \p_V.
\label{MinkSL}
\eeq
The fifth isometry is the ``vertical translation" generated by $\p_V$.

Things become more complicated if the profile is non-trivial. In the linearly polarized case  $\cB=0$ with a time independent profile $\cA=D\neq0$ a real constant, for example,
transverse translations are no longer isometries.
In this simple but rather non-physical (non-sandwich) case
Eqn. (\ref{SLeqn}) decouples into two time-independent equations of the oscillator-form, one of them attractive and the other repulsive, depending on the sign of $D$. 
The solution of (\ref{SLeqn}) is therefore a 4-parameter  linear combinations  which mixes $\sinh( \sqrt{|D|}\,U) $ and
  $\cosh( \sqrt{|D|}\,U)$ in the repulsive, and
  $\sin (\sqrt{|D|}\,U)$ and $\cos(\sqrt{|D|}\,U)$ in the attractive component.
But when $D\neq\const$, the Sturm-Liouville problem (\ref{SLeqn})
 has no analytic solution in general, and in the polarized case  with time dependent profile everything becomes even worse. 
Various properties of the Killing vector fields and the group they generate were explored  in  \cite{EhEm}.

%%%%%%%%%%%%%%%%%%%%%%%%%%%%%%%%%%%%%%%%
%\subsection{In Baldwin-Jeffery-Rosen coordinates}\label{RosenSec}
%%%%%%%%%%%%%%%%%%%%%%%%%%%%%%%%%%%%%%%%

The problem of solving the Sturm-Liouville equation (\ref{SLeqn}) was circumvented by Souriau \cite{Sou73} who suggested using  BJR coordinates \cite{BaJe} instead, in terms of which the metric may be written as
\beq
a_{ij}(u)\,dx^idx^j+2du\,dv,
\label{BJRmetrics}
\eeq
with $a(u)=(a_{ij}(u))$ a positive definite $2\times2$ matrix, which is an otherwise arbitrary function of ``non-relativistic time'', $u$ \footnote{The BJR coordinates are valid only in finite intervals before becoming singular \cite{BoPi89}, see sec. \ref{BRosenSec}.}. Then natural  translations 
$\bx \to \bx+\bc,\, v\to v+ w$ are manifest isometries. Moreover, implementing  $\bb\in\bbR^2$ through the $2\times2$ matrix valued function
\beq
H(u)=\int_{u_0}^u{\!a^{-1}(w)dw}\,,
\label{Hmatrix}
\eeq\vspace{-4mm}
as \cite{Sou73,Carroll4GW},\vspace{-4mm}
\begin{subequations}
\begin{align}
\label{Carrollx}
\bx&\to\bx+ H(u)\,\bb,
\\
u&\to u,
\\
v&\to v-\bb\cdot\bx - \2\bb\cdot{}H(u)\,\bb,
\label{Carrollv}
\end{align}
\label{genCarr}
\end{subequations}
yields two further isometries.  
Souriau's  form (\ref{genCarr}) allows one to determine the group structure with the remarkable result that the above-mentioned generic $5$-parameter isometry group \cite{BoPiRo,exactsol} we denote here by $G$ may  be identified, \emph{for any profile}, as \emph{the subgroup of the Carroll group
with rotations omitted} \cite{Carroll4GW}.
%%%.  

The transformations (\ref{genCarr}) generated by  $\bb$ are, in particular, \emph{boosts}, implemented in an unusual way. In the Minkowski case
$a=\Id$ is the unit matrix so that
$ 
H_{Mink}(u)=(u-{u_0})\,\Id\,
$ 
and the standard extension of the Galilei group  [lifted to  flat Bargmann space] is recovered.

%%%%%%%%%%%%%%%%%%%%%%%%%%%%%%%%%%%%%%%%%%%%
\section{Brinkmann $\Leftrightarrow$ BJR}\label{BRosenSec}
%%%%%%%%%%%%%%%%%%%%%%%%%%%%%%%%%%%%%%%%%%%%

The relation between Brinkmann and BJR coordinates  is 
\cite{Gibb75}
\beq
{\bX}=P(u)\,\bx,
\quad
U=u,
\quad 
V=v-\frac{1}{4}\bx\cdot\dot{a}(u)\bx
\quad\text{with}\quad
a(u)=P^{T}\!(u){}P(u)\,,
\label{BBJRtrans}
\eeq
where the $2\times2$ matrix $P$ satisfies
\beq
\ddot{P}=K\,P\,,
\qquad
P^{T}\dot{P}-\dot{P^{T}}P=0\,.
\label{SLP}
\eeq
%%%%%%%%%
The first of these is  a \emph{matrix  Sturm-Liouville equation} for $P$. 
If this is satisfied, then $P^{T}\dot{P}-\dot{P^{T}}P=0$ is  shown to be a constant of the motion. The second equation is therefore satisfied when it holds at an arbitrary moment.

We emphasise that while the B-coordinates are global, the BJR coordinates are valid  only in a finite interval: the mapping
(\ref{BBJRtrans}) necessarily becomes singular when 
\beq
\det (a)=0\,
\qquad
\text{\small or equivalently}
\qquad
\det(P)=0\,.
\label{caustic}
\eeq 
%%%% 
The mapping (\ref{BBJRtrans}) trades the quadratic ``potential''  $K_{ij}(U){X^i}{X^j}$ in (\ref{genBrink}) for a ``time''-dependent transverse metric $a(u)=\big(a_{ij}(u)\big)$ in (\ref{BJRmetrics}) and vice versa.

%%%%%%%%%%%% 
When expressed in B coordinates, the natural BJR transverse translations $\bx \to \bx +\bc$ become
``time-dependent translations of the Newton-Hooke form'' \cite{GiPo}
\beq
\bX \to \bX + P(u)\,\bc\,.
\label{Btransl}
\eeq
%%%%%%
Further insight is gained by introducing the $2\times2$ matrix 
\beq
Q(U)=
\big(P{H}\big)(U)=P(U)\int_{{U_0}}^U\big(P^{-1}(P^T)^{-1}\big)(w)dw\;.
\label{BinB}
\eeq
Then combining (\ref{genCarr}) with (\ref{BBJRtrans}) yields, for boosts,
%%%%%%%%%%%%
\besub
\begin{align}
&\bX \to  \bX+ Q\, \bb\,,
\label{BXsymimp}
\\
& V \to V 
- \bX\cdot\dot{Q}\,\bb   
-\frac{1}{2} Q\,\bb\cdot \dot{Q}\,\bb\,.
\label{BVsymimp}
\end{align}
\label{Bsymimp}
\esub
Eqn. (\ref{Bsymimp}) can be tested by
inserting the Minkowskian values; reassuringly, the usual Galilean expression is recovered.
Moreover, a straightforward calculation using (\ref{Hmatrix}), (\ref{BBJRtrans}) and (\ref{SLP}) shows that 
$Q$ satisfies the same matrix Sturm-Liouville equations (\ref{SLP}) as $P$ does,  
\beq
\ddot{Q}=K\,Q\,,
\qquad
Q^{T}\dot{Q}-\dot{Q^{T}}Q=0.
\label{SLS}
\eeq

Working infinitesimally, only those terms which are linear in $\bb$ contribute, and we recover 
 the rule (\ref{TorreSymm}) of Torre \cite{Torre}
with $S_i=P\bc_i,\, \bc_1=(1,0)$ 
and $\bc_2=(0,1)$ for translations, and
$S_i=Q\bb_i$ for boosts, respectively.

%%%%%%%%%%%%%%%%%%%%%%%%%%%%%%%%%% 
\section{Trajectories}\label{TrajSec}
%%%%%%%%%%%%%%%%%%%%%%%%%%%%%%%%%%  

%%%%%%%%%%%%%%%%%%%%%%%%%%%%
%\subsection{Sandwich waves}
%%%%%%%%%%%%%%%%%%%%%%%%%%%%

 The $G$-symmetry implies that \footnote{The conserved quantity associated with the ``vertical'' Killing vector
$\p_V$ can be used to show that proper time and $u$ are proportional.}
\beq
\bp\!=\! a(u)\,\dot{\bx},
\qquad
\bk\!=\!\bx(u)-H(u)\,\bp,
\label{CarCons}
\eeq
 interpreted as conserved \emph{linear and  boost-momenta}.  
%%%%%
Reversing these relations, 
the geodesics may be expressed  using the Noetherian quantities above \cite{Sou73,ShortLong}, 
\begin{equation}
\bx(u)=H(u)\,\bp+\bk,
\qquad
v(u)=-\half \bp\cdot H(u)\,\bp + e\,u+ v_0,
\label{BJRGeo}
\end{equation}\vskip-2mm\noindent
where \vspace{-3mm}
\beq 
e=\half g_{\mu\nu}\dot{x}^\mu\dot{x}^\nu
\label{Jacobi}
\eeq
is another constant of the motion, whose sign only
depends on the nature (timelike/spacelike or null) of the geodesic. $v_0$ is a constant of integration.
Once the values of the conserved quantities are fixed, the only quantity to be calculated is 
 the matrix-valued function $H(u)$ in (\ref{Hmatrix}), which
 thus determines both how the isometries act, (\ref{genCarr}),  and also the evolution of causal geodesics, (\ref{BJRGeo}). 
$H(u)$ is in turn related to the Sturm-Liouville solution $P$ in (\ref{SLP}).  In flat Minkowski space 
 $H(u)=u\,\Id$ yields free motion, 
$ 
\bx(u)=(u-u_0)\,\bp+\bk,\, 
v(u)= (u-u_0)\left(-\half {\vert\bp\vert}^2 +e\right)+ v_0.
$ 

Returning to the general metric (\ref{BJRmetrics}), requiring that our particles be at rest before the gravitational wave arrives implies,  by (\ref{CarCons}), $\bp=0$ for all profiles~: 
\beq
\bx=\bx_0 \equiv \bk=\const, \qquad v=(u-{u_0})e + v_0.
\label{simpleBJRtraj}
\eeq
With tongue-in-cheek, we call it ``Carrollian'', since there is no (transverse) motion ---  the hallmark of ``Carrollian'' physics \cite{Leblond,Ancille,Carrollvs,Bergshoeff} \footnote{Remember that the 4D gravitational wave spacetime is in fact the ``Bargmann space'' for both a non-relativistic and for a Carroll particle in the transverse plane \cite{Bargmann,Carrollvs}; geodesics in $4D$ project to motions in $2+1$ dimensions.}.

%%%%%%%%%%%%%%%%%%%%%%%%%%%%%%%%%%%%%%%%%%%%%%
%\subsection{Carroll action \& straightening out}
%%%%%%%%%%%%%%%%%%%%%%%%%%%%%%%%%%%%%%%%%%%%%%
 
Remarkably, \emph{every} geodesic is obtained from a simple one of the form (\ref{simpleBJRtraj}) by a suitable symmetry transformation \cite{EhlersKundt,Sou73,Carroll4GW}.
Eqn. (\ref{BJRGeo}) says indeed that any geodesic determines and is, conversely, determined by six constants of the motion. Then we note that the isometry group $G$
acts on the constants of the motion $\Phi=(\bp,\bk,e,v_0)$ according to \cite{Sou73,Carroll4GW}
\beq
(\bp,\bk,e,v_0) \to (\bp+\bb,\bk+\bc,e, v_0+f-\bb\cdot\bk).
\label{Carrolloncq}
\eeq
This action leaves  $e$ in (\ref{Jacobi}) invariant, however for any fixed value of $e$ we can find an appropriate element of $G$
 which  brings $\Phi=(\bp,\bk, e, v_0)$ to  $\Phi_0=(\bp=0,\bk=0, e, v_0=0)$. Conversely, any given set $\Phi$ can be reached from $\Phi_0$ by the action of an isometry.
The geodesic with parameters $\Phi_0$ is
(\ref{simpleBJRtraj}) with $x_0=0$ and $v_0=0$, for which the trajectory is simply
$ 
\bx(u)=0\,,\; v(u) =  (u-{u_0})\,e.
$ 
Conversely, the geodesic with parameters $\Phi$ is obtained in turn by implementing the $\Phi_0\to\Phi$ isometry as in (\ref{genCarr}), as illustrated on Fig.3 of \cite{Carroll4GW}.

One can wonder how all this looks in B coordinates. The answer can be obtained by ``exporting'' the trajectories from BJR to B using (\ref{BBJRtrans}) \footnote{  
Comparison with the trajectories obtained  by solving directly the equations of motion numerically shows a perfect overlapping.
This is a third appearance of the solution $P$ of the SL eqn (\ref{SLP}). In Souriau's approach it is the determinant of the metric (\ref{genBrink}) which satisfies a Sturm-Liouville equation.}. Choosing $\bX_0=\bx_0$ in the before zone,
\besub\vspace{-5mm}
\begin{align}
&\bX(U)= P(U)\,\bX_0, 
\label{BXsimplegeo}
\\[2pt]
&V(U) =  
(U-{U}_0)e+v_0 - \dfrac{1}{4}\frac{d(\bX^2)}{dU}(U)
\label{BVsimplegeo}
\end{align}
\label{Bsimplegeo}
\esub
which, for $\bX_0=0$ and $v_0=0$, remains ``Carrollian'' in that  
$ 
\bX(U)\!=\!0,\, V(U)\!= \!(U-{U_0})\,e.
$

%%%%%%%%%%%%%%%%%%%%%%%%%%%%%%%%%%%%%%%%%%%%%%%
\section{Illustration: polarized sandwich waves}
%%%%%%%%%%%%%%%%%%%%%%%%%%%%%%%%%%%%%%%%%%%%%%%

\begin{figure}[h]
\includegraphics[scale=.26]{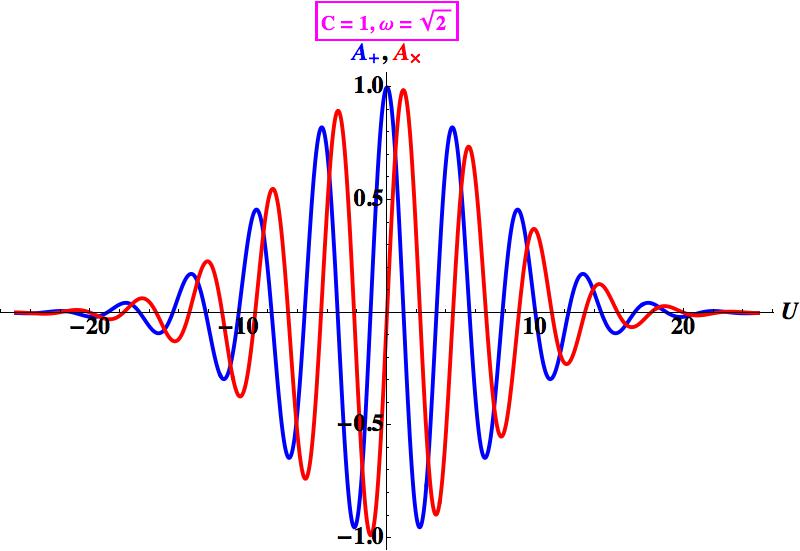}\\
\null\vskip-11mm
\caption{\textit{\small  Polarized sandwich wave with Gaussian envelope as given in (\ref{polgaussprof}).
The colors refer to the $\blue{\cA}$ and the \red{$\cB$} polarisation components.
}}
\vskip-3mm
\label{SandProf}
\end{figure}

Let us indeed consider  a circularly polarised oscillating  sandwich wave with Gaussian envelope (\ref{polgaussprof}), shown in Fig. \ref{SandProf}.
 The simple $\bp=0$ trajectory (\ref{simpleBJRtraj})  describes the motion of a particle at rest in the before zone.
 BJR coordinates are valid between caustics, which appear where $\det\big(P(u_1)\big)=0$; numerically, we found, in the neighborhood of $u_0=0$, $u_1= -2.80< u <  u_2 = 2.74$; Brinkmann coordinates work for all $U$.

\begin{figure}[h]
\bigskip
\includegraphics[scale=.26]{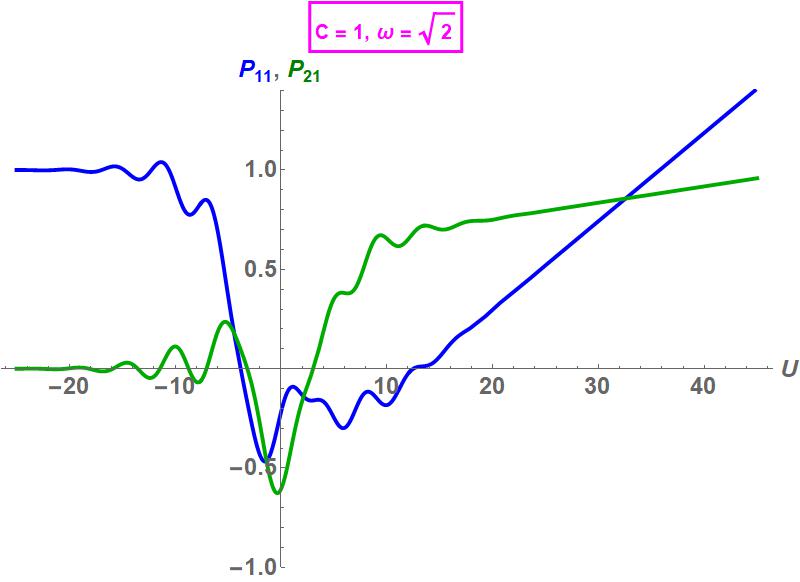}
\;\;\;
\includegraphics[scale=.26]{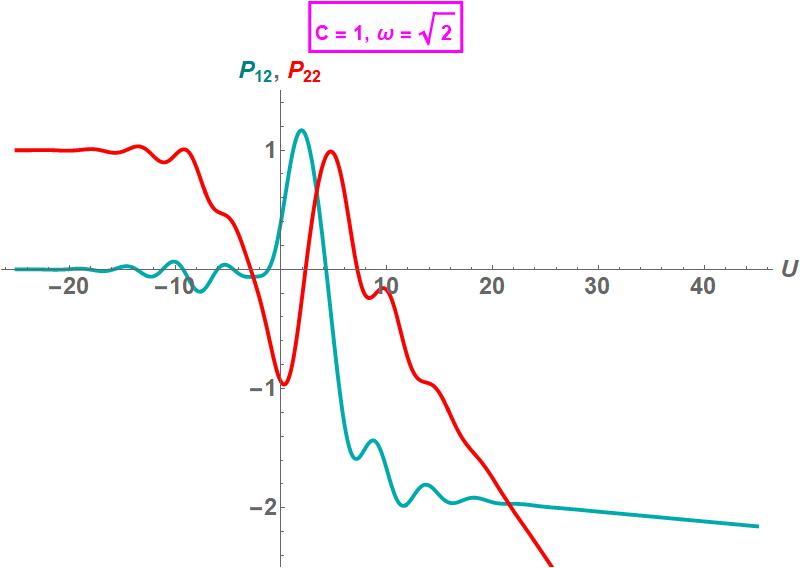}
\\
\null\vskip-11mm
\caption{\textit{\small The images under the B $\Leftrightarrow$ BJR map (\ref{Bsimplegeo}) of the simple trajectories  (\ref{simpleBJRtraj})  initially at rest for $u_0<0$ at $\bx_0^1=\bX_0^{(1)}=(\blue{1},\dgreen{0})$ and at
$\bx_0^{(2)}=\bX_0^{(2)}=(\red{0},\cyan{1})$, respectively, are, B coordinates, the two columns of the $P$ matrix. The motion is complicated in the inside-zone but follows straight lines with constant velocity in the after-zone.
}}
%\vskip-6mm
\label{BstraightGeo}
\end{figure}

Then  (\ref{Bsimplegeo}) can be plotted after solving the Sturm-Liouville eqn (\ref{SLP}) numerically.  
Fig.\ref{BstraightGeo} shows, however, that even such simple trajectories become complicated-looking, with the exception of the one for $\bX_0=0$.
%
%\vskip-6mm
%%%%%%%%%%%%%%%%%

%\vskip-6mm
%%%%%%%%%%%%%%%%%
\begin{figure}[h]
\includegraphics[scale=.27]{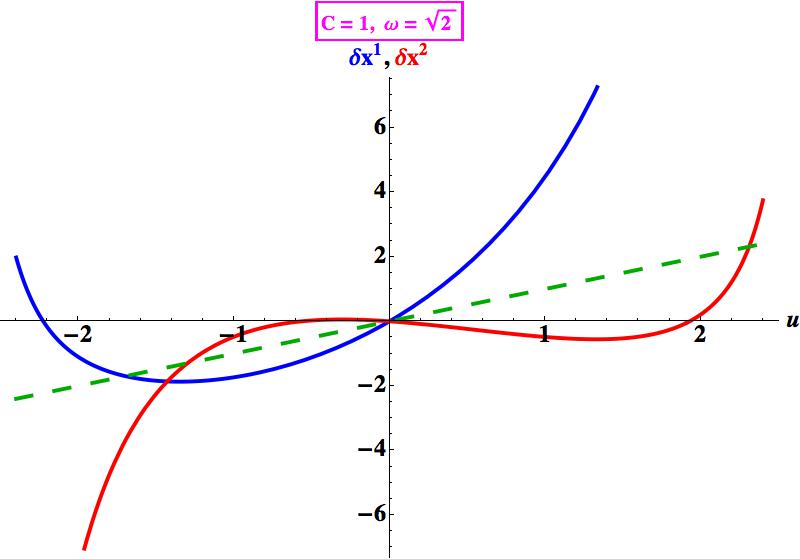}
\null\vskip-5mm
\caption{\textit{\small In BJR coordinates
boosts act according to (\ref{genCarr}). The  implementation differs substantially from the \dgreen{Galilean} one (\dgreen{dashed}). We took here $\bb=(1,1)$.  
}}
\label{impHfig}
\end{figure}

\goodbreak
%
%%%%%%%%%%%%%%%%%%%%%%%%%%%
Having calculated (numerically) the matrix  $P$, we can proceed to calculating $a=P^T\!P$ and then $H$ in (\ref{Hmatrix}), allowing us to plot 
finally how boosts are implemented in BJR coordinates in a neighborhood of the origin, $\bx \to \bx + \delta\bx,\, \delta \bx= H(u)\,\bb$, cf.  (\ref{genCarr}), shown on Fig.\ref{impHfig}. The  implementation differs substantially from the Galilean one, $\delta \bx = u\bb$.
%%%%%%%%%%%%%%%%%

\vskip-3mm
%%%%%%%%%%%%%%%%%%%%%%%%%%%%%%%%%%%%
\section{Conclusion}
%%%%%%%%%%%%%%%%%%%%%%%%%%%%%%%%%%%%

The Memory Effect boils down to solving the Sturm-Liouville equation (\ref{SLPeq}) -- a task which can, in general, be done only numerically. 

Particles at rest in the before zone have vanishing momenta, implying that in BJR coordinates the trajectory is trivial~: all those complicated-looking trajectories obtained before \cite{ShortLong,ImpMemory,PolPer} are in fact images of the trivial ``Carrollian'' ones in (\ref{simpleBJRtraj}) resp. (\ref{BVsimplegeo}) by a suitable broken-Carroll isometry \cite{EhlersKundt,Sou73,Carroll4GW}~: all complications are hidden in the Sturm-Liouville equation (\ref{SLPeq}). 
This can be viewed as the gravitational-wave generalization of the observation that \emph{any free non-relativistic motion is obtained from static equilibrium by a Galilei transformation}. The ``no motion'' defect of Carrollian dynamics \cite{Leblond,Ancille,Carrollvs,Bergshoeff} is thus turned into an advantage.
\vskip-2mm

%\%\%\%
\begin{acknowledgments} 
We are grateful to Christian Duval for his contribution at the early stages of this project, and 
to an anonymous referee for drawing our attention to
\cite{EhEm} of which were were previously unaware.
 ME and  PH  thank the \emph{Institute of Modern Physics} of the Chinese Academy of Sciences in Lanzhou for hospitality.  This work was supported by the Chinese Academy of Sciences President's International Fellowship Initiative (No. 2017PM0045), and by the National Natural Science Foundation of China (Grant No. 11575254).
 PH would like to acknowledge also the organizers of the \textit{``Workshop on Applied Newton-Cartan Geometry''}  and the \textit{Mainz Institute for Theoretical Physics} (MITP), where part of this work was completed. We are grateful to our colleagues
to inform us about their work in progress \cite{AndrPrenc}.
\end{acknowledgments}
\goodbreak

%%%%%%%%%%%%%%%%%%%%%%%%%%%%%%%%%%%%%%%%%%%%%%%%%%%%%%%%%%%%%%%%%%%%%%%%%%%%%%
%%%%%%%%%%%%%%%%%%%%%%%%%%%%%%%%%%%%%%%%%%%%%%%%%%%%%%%%%%%%%%%%%%%%%%%%%%%%%%

%%%%%%%%%%%%%%

\begin{thebibliography}{99} 
%%%%%%%%%%%%%%%%%%%%%%%%%%%%%%%%%%%%%%%%%%%%%%%%%%%%%%%%%%%%%%%%%%%%%%%%%%%%%%
%%%%%%%%%%%%%%%%%%%%%%%%%%%%%%%%%%%%%%%%%%%%%%%%%%%%%%%%%%%%%%%%%%%%%%%%%%%%%%

\bibitem{ZelPol}
Ya. B. Zel'dovich and A. G. Polnarev,
``Radiation of gravitational waves by a cluster of superdense stars,"
Astron. Zh. {\bf 51}, 30 (1974)
[Sov. Astron. {\bf 18} 17 (1974)].

\bibitem{BraGri}
 V B Braginsky and L P Grishchuk,
``Kinematic resonance and the memory effect in
free mass gravitational antennas,''
 Zh. Eksp. Teor. Fiz. {\bf 89} 744-750 (1985) [Sov. Phys. JETP 62, 427 (1985)].
   
\bibitem{EhlersKundt}
%\cite{Ehlers:1962zz}
  J.~Ehlers and W.~Kundt,
 ``Exact solutions of the gravitational field equations,'' 
 in L. Witten, (ed), {\it Gravitation, an introduction to current Research}, Wiley, New York, London, (1962).
  %%CITATION = INSPIRE-45501;%% 
   
\bibitem{Sou73}
J-M. Souriau,
``Le milieu \'elastique soumis aux ondes gravitationnelles,''  \textit{Ondes et radiations gravitationnelles}, 
Colloques Internationaux du CNRS No 220, p. 243. Paris (1973).  
     
\bibitem{BraTho}
 V. B. Braginsky and  K. S. Thorne,
``Gravitational-wave burst with memory and experimental prospects,''
 Nature (London) {\bf 327} 123 (1987). 

%\cite{Bondi:1989vm}
\bibitem{BoPi89}
  H.~Bondi and F.~A.~E.~Pirani,
  ``Gravitational Waves in General Relativity. 13: Caustic Property of Plane Waves,''
  Proc.\ Roy.\ Soc.\ Lond.\ A {\bf 421} (1989) 395.
 % doi:10.1098/rspa.1989.0016
  %%CITATION = doi:10.1098/rspa.1989.0016;%%
  
\bibitem{GriPol}
%\cite{Grishchuk:1989qa}
%\bibitem{Grishchuk:1989qa}
  L.~P.~Grishchuk and A.~G.~Polnarev,
  ``Gravitational wave pulses with `velocity coded memory',''
  Sov.\ Phys.\ JETP {\bf 69} (1989) 653
   [Zh.\ Eksp.\ Teor.\ Fiz.\  {\bf 96} (1989) 1153].
  %%CITATION = SPHJA,69,653;%%
   
\bibitem{ShortLong}
%\cite{Zhang:2017rno}
%\bibitem{ShortMemory}
  P.-M.~Zhang, C.~Duval, G.~W.~Gibbons and P.~A.~Horvathy,
   ``The Memory Effect for Plane Gravitational Waves,''
  Phys.\ Lett.\ B {\bf 772} (2017) 743.
  doi:10.1016/j.physletb.2017.07.050
  {}[arXiv:1704.05997 [gr-qc]].
  %%CITATION = doi:10.1016/j.physletb.2017.07.050;%%
%\bibitem{LongMemory}
%\cite{Zhang:2017geq}
%\bibitem{Zhang:2017geq}
%  P.-M.~Zhang, C.~Duval, G.~W.~Gibbons and P.~A.~Horvathy,
  ``Soft gravitons and the memory effect for plane gravitational waves,''
  Phys.\ Rev.\ D {\bf 96} (2017) no.6,  064013
  doi:10.1103/PhysRevD.96.064013.
  [arXiv:1705.01378 [gr-qc]].
  %%CITATION = doi:10.1103/PhysRevD.96.064013;%% 

\bibitem{Lasenby}
A. Lasenby, ``Black holes and gravitational waves,''
talks  given at the Royal Society Workshop on `Black Holes', Chichley Hall, UK (2017) and KIAA, Beijing (2017).
   

  %\cite{Zhang:2018srn}
\bibitem{PolPer}
%\cite{Zhang:2018srn}
%\bibitem{Zhang:2018srn}
  P.~M.~Zhang, C.~Duval, G.~W.~Gibbons and P.~A.~Horvathy,
  ``Velocity Memory Effect for Polarized Gravitational Waves,''
  JCAP {\bf 1805} (2018) no.05,  030
  doi:10.1088/1475-7516/2018/05/030
  [{\tt arXiv:1802.09061 [gr-qc]}].
  %%CITATION = doi:10.1088/1475-7516/2018/05/030;%%
 
%\cite{Gibbons:1975jb}
\bibitem{Gibb75}
 G.~W.~Gibbons,
  ``Quantized Fields Propagating in Plane Wave Space-Times,''
  Commun.\ Math.\ Phys.\  {\bf 45} (1975) 191.
% doi:10.1007/BF01629249  
  
\bibitem{Leblond} 
J. M. L\'evy-Leblond, 
``Une nouvelle limite non-relativiste du group de Poincar\'e,''
Ann. Inst. H Poincar\'e {\bf 3} (1965) 1; %-12;
%\bibitem{SenGupta}
 V. D. Sen Gupta, 
 ``On an Analogue of the Galileo Group,'' 
 Il Nuovo Cimento {\bf 54} (1966) 512. %-517 
 
%\cite{Bondi:1958aj}
\bibitem{BoPiRo}
  H.~Bondi, F.~A.~E.~Pirani and I.~Robinson,
 ``Gravitational waves in general relativity. 3. Exact plane waves,''
  Proc.\ Roy.\ Soc.\ Lond.\ A {\bf 251} (1959) 519.
  doi:10.1098/rspa.1959.0124
  %%CITATION = doi:10.1098/rspa.1959.0124;%%
  
\bibitem{exactsol}
D. Kramer, H. Stephani, M. McCallum, E. Herlt,
``Exact solutions of Einstein's field equations,''
 Cambridge Univ. Press 2nd ed. (2003) sec 24.5 Table 24.2, p.385.  

%\cite{Duval:2017els}
\bibitem{Carroll4GW}
 C.~Duval, G.~W.~Gibbons, P.~A.~Horvathy and P.-M.~Zhang,
``Carroll symmetry of plane gravitational waves,''
Class. Quant. Grav. {\bf 34} (2017).
doi.org/10.1088/1361-6382/aa7f62. [arXiv:1702.08284 [gr-qc]].

\bibitem{Ancille}
%\cite{Ngendakumana:2013hza}
%\bibitem{Ngendakumana:2013hza}
  A.~Ngendakumana, J.~Nzotungicimpaye and L.~Todjihounde,
  ``Group theoretical construction of planar Noncommutative Phase Spaces,''
  J. Math. Phys. {\bf 55}, 013508 (2014) %
  [arXiv:1308.3065 [math-ph]].  
  
%\cite{Duval:2014uoa}
\bibitem{Carrollvs}
  C.~Duval, G.~W.~Gibbons, P.~A.~Horvathy and P.~M.~Zhang,
  ``Carroll versus Newton and Galilei: two dual non-Einsteinian concepts of time,'' %\\
{\it Class. Quant. Grav.} {\bf 31} (2014) 085016
 [arXiv:1402.5894 [gr-qc]].
 %%CITATION = ARXIV:1402.0657;%%   
  
%\cite{Bergshoeff:2014jla}
\bibitem{Bergshoeff}
  E.~Bergshoeff, J.~Gomis and G.~Longhi,
  ``Dynamics of Carroll Particles,''
  Class.\ Quant.\ Grav.\  {\bf 31} (2014) no.20,  205009
  doi:10.1088/0264-9381/31/20/205009
  [arXiv:1405.2264 [hep-th]].
  %%CITATION = doi:10.1088/0264-9381/31/20/205009;% 
 
\bibitem{Brink}
M. W. Brinkmann,
 ``Einstein spaces which are mapped conformally on each other,''
 Math. Ann. {\bf 94} (1925)~119--145.
  
\bibitem{ImpMemory}
%\cite{Zhang:2017jma}
%\bibitem{Zhang:2017jma}
  P.-M.~Zhang, C.~Duval and P.~A.~Horvathy,
 ``Memory Effect for Impulsive Gravitational Waves,''
  Class.\ Quant.\ Grav.\  {\bf 35} (2018) no.6,  065011
   doi:10.1088/1361-6382/aaa987
  [arXiv:1709.02299 [gr-qc]].
  %%CITATION = doi:10.1088/1361-6382/aaa987;%%   
 
%\cite{Podolsky:2014ysa}
\bibitem{PodSB}
  J.~Podolsk\'y, C.~S\"amann, R.~Steinbauer and R.~Svarc,
  ``The global existence, uniqueness and $C^1$-regularity of geodesics in nonexpanding impulsive gravitational waves,''
  Class.\ Quant.\ Grav.\  {\bf 32} (2015) no.2,  025003
  doi:10.1088/0264-9381/32/2/025003
  [arXiv:1409.1782 [gr-qc]].
  %%CITATION = doi:10.1088/0264-9381/32/2/025003;%%

%\cite{Torre:1999ye} 
\bibitem{Torre}
C.~G.~Torre,
``Gravitational waves: Just plane symmetry,''
Gen.\ Rel.\ Grav.\  {\bf 38} (2006) 653
%�doi:10.1007/s10714-006-0255-8
[gr-qc/9907089].

\bibitem{Bargmann}
%\cite{Duval:1984cj}
%\bibitem{Duval:1984cj}
C. Duval, G. Burdet, H. P. K\"{u}nzle and M. Perrin,
 ``Bargmann structures and Newton-Cartan theory",
Phys. Rev. D {\bf 31} (1985) 1841;

\bibitem{DGH91}
C. Duval, G.W. Gibbons, P. Horvathy,
 ``Celestial mechanics, conformal structures and gravitational waves,''
 Phys. Rev. {\bf D43} (1991) 3907. %-3922
 [hep-th/0512188].
 
 \bibitem{EhEm} 
%\cite{Ehrlich:1992fe}
%\bibitem{Ehrlich:1992fe}
  P.~E.~Ehrlich and G.~G.~Emch,
 ``Gravitational waves and causality,''
  Rev.\ Math.\ Phys.\  {\bf 4} (1992) 163
   Erratum: [Rev.\ Math.\ Phys.\  {\bf 4} (1992) 501].
  doi:10.1142/S0129055X92000066
  %%CITATION = doi:10.1142/S0129055X92000066;%%

\bibitem{BaJe}
O. R. Baldwin and G. B. Jeffery,
``The Relativity Theory of Plane Waves,''
Proc. R. Soc. London {\bf A111}, 95 (1926).
% \bibitem{Ros}
 N. Rosen,
``Plane polarized waves in the general theory of relativity,''
Phys. Z. Sowjetunion, {\bf 12}, 366 (1937). See  also
%\bibitem{LaLi}
L.D. Landau \& E.M. Lifshitz,
``The Classical Theory of Fields,'' 
(Volume 2 of \textit{A Course of Theoretical Physics}) 
Pergamon Press (1971). 

\goodbreak
%\cite{Gibbons:2010fb}
\bibitem{GiPo}
  G.~W.~Gibbons and C.~N.~Pope,
  ``Kohn's Theorem, Larmor's Equivalence Principle and the Newton-Hooke Group,''
  Annals Phys.\  {\bf 326} (2011) 1760
  doi:10.1016/j.aop.2011.03.003
  [arXiv:1010.2455 [hep-th]];
  %%CITATION = doi:10.1016/j.aop.2011.03.003;%%
%\cite{Zhang:2012cr}
%\bibitem{Zhang:2012cr}
 P.~M.~Zhang, P.~A.~Horvathy, K.~Andrzejewski, J.~Gonera and P.~Kosinski,
``Newton-Hooke type symmetry of anisotropic oscillators,''
 Annals Phys.\  {\bf 333} (2013) 335
  [arXiv:1207.2875 [hep-th]].
  %%CITATION = ARXIV:1207.2875;%%

\bibitem{AndrPrenc}
%\cite{Andrzejewski:2018pwq}
%\bibitem{Andrzejewski:2018pwq}
  K.~Andrzejewski and S.~Prencel,
  ``Memory effect, conformal symmetry and gravitational plane waves,''
  Phys.\ Lett.\ B {\bf 782} (2018) 421
  doi:10.1016/j.physletb.2018.05.072
  [arXiv:1804.10979 [gr-qc]].
  %%CITATION = doi:10.1016/j.physletb.2018.05.072;%%
 
\end{thebibliography}
\end{document}